# General Microscopic Model of Magnetoelastic Coupling from First-Principles

X. Z. Lu[1], Xifan Wu[2], and H. J. Xiang[1,3]*

[1]Key Laboratory of Computational Physical Sciences (Ministry of Education), State Key Laboratory of Surface Physics, and Department of Physics, Fudan University, Shanghai 200433, P. R. China

[2]Department of Physics, Temple Materials Institute, and Institute for Computational Molecular Science, Temple University, Philadelphia, Pennsylvania 19122, USA

[3]Collaborative Innovation Center of Advanced Microstructures, Fudan University, Shanghai 200433, China

e-mail: hxiang@fudan.edu.cn

**Abstract**

Magnetoelastic coupling, i.e., the change of crystal lattice induced by a spin order, is not only scientifically interesting, but also technically important. In this work, we propose a general microscopic model from first-principles calculations to describe the magnetoelastic coupling and provide a way to construct the microscopic model from density functional theory calculations. Based on this model, we reveal that there exists a previously unexpected contribution to the electric polarization induced by the spin-order in multiferroics due to the combined effects of magnetoelastic coupling and piezoelectric effect. Interestingly and surprisingly, we find that this lattice deformation contribution to the polarization is even larger than that from the pure electronic and ion-displacement contributions in $BiFeO_3$. This model of magnetoelastic coupling can be generally applied to investigate the other magnetoelastic phenomena.

Magnetoelasticity refers to the phenomenon where a change of magnetic state can induce a change in crystal volume/shape and vice versa. The study of this phenomenon can be traced back to 1960s [1,2]. Magnetoelastic materials are playing an increasingly important role in applications ranging from actuation, sensing, and energy harvesting [3]. The large scientific interest in the magnetoelastic coupling is connected to its fundamental importance in many research areas. For example, in some negative thermal expansion (NTE) magnetic material [4-8], the system shows abrupt increase in crystal volume on cooling in the vicinity of the magnetic transition from the paramagnetic (PM) state to ordered magnetic state. In some frustrated spin systems such as spinel $ACr_2O_4$ (A=Mg, Zn) [9-12], the magnetoelastic coupling causes a change of the crystal lattice from cubic to tetragonal when they undergo an antiferromagnetic (AFM) phase transition. Furthermore, in the phenomenon of magnetostriction [3], the strain dependence of the magnetic anisotropy and/or exchange interactions can lead to a lattice change in the certain direction when a magnetic field is applied. First-principles density function theory (DFT) calculations [13-15] have been performed to understand magnetoelasticity (in particularly magnetostriction). While direct DFT calculations agree well the macroscopic lattice response associated with various magnetic configurations, a theoretical model that elucidates the microscopic origin will be desired.

For dielectric materials, the response properties can be systematically treated by electric-magnetic enthalpy as functions of ionic displacement, strain, applied electric, and magnetic fields [16,17]. Here in this paper, we further develop a first-principles based model describing magnetoelastic coupling. In this model, the relationship between the change of crystal lattice and spin order is simplified to two linear equations from which the atomic displacements and strains induced by the spin order can be obtained simultaneously, thus quantitatively describing the lattice changes. This model is general so that it can be adopted to understand the other magnetoelastic related phenomena [including symmetric exchange, antisymmetric Dzyaloshinskii-Moriya (DM) interaction and single-ion anisotropy (SIA) related cases]. According to our model, we reveal that there is a new contribution (i.e., lattice deformation) to the spin-order induced electric polarization in multiferroics: The spin order induces a lattice strain, which subsequently gives rise to an additional electric polarization through the piezoelectric effect [16,18]. By combining our model with DFT calculations, we demonstrate that the lattice deformation contribution is larger than the pure electronic and ionic contributions in BiFeO$_3$.

In general, the total energy of a localized magnetic system can be written as E($u_m$, $\eta_j$, $\mathbf{S}_i$)=E$_{PM}$($u_m$, $\eta_j$)+E$_{spin}$($u_m$, $\eta_j$, $\mathbf{S}_i$), where $u_m$ is the atomic displacement from a reference structure, $\eta_j$ (j ={1…6}) is the homogeneous strain in Voigt notation, and $\mathbf{S}_i$ refers to the spin vector. Here, E$_{PM}$ is the energy of the paramagnetic (PM) state which can be expanded as [16, 17]:

$$E_{PM} = E_0 + A_m u_m + A_j \eta_j + \frac{1}{2} B_{mn} u_m u_n + \frac{1}{2} B_{jk} \eta_j \eta_k + B_{mj} u_m \eta_j \quad (1)$$
+terms of third and higher orders.

The first-order coefficients A$_m$ and A$_j$ and the second-order coefficients B$_{mn}$, B$_{jk}$ and B$_{mj}$ represent force, stress, force constant, frozen-ion elastic constant, and internal-displacement tensor, respectively. By choosing a reference structure that is in equilibrium in the PM state, we will have A$_m$=A$_j$=0. It should be noted that an implied-sum notation is adopted in this work. The spin interaction energy E$_{spin}$ usually contains three parts [12] (E$_{spin}$ = E$_H$ + E$_{DM}$ + E$_{SIA}$): the Heisenberg symmetric exchange interaction E$_H$, antisymmetric Dzyaloshinskii-Moriya (DM) interaction E$_{DM}$, and single-ion anisotropy (SIA) E$_{SIA}$. The Heisenberg exchange interaction E$_H$ can be expanded as:

$$E_H = E_H^0 + \sum_{i,i'} \frac{\partial J_{ii'}}{\partial u_m} \mathbf{S}_i \cdot \mathbf{S}_{i'} u_m + \sum_{i,i'} \frac{\partial J_{ii'}}{\partial \eta_j} \mathbf{S}_i \cdot \mathbf{S}_{i'} \eta_j + \sum_{i,i'} \frac{\partial^2 J_{ii'}}{\partial u_m \partial u_n} \mathbf{S}_i \cdot \mathbf{S}_{i'} u_m u_n$$
$$+ \sum_{i,i'} \frac{\partial^2 J_{ii'}}{\partial \eta_j \partial \eta_k} \mathbf{S}_i \cdot \mathbf{S}_{i'} \eta_j \eta_k + \sum_{i,i'} \frac{\partial^2 J_{ii'}}{\partial u_m \partial \eta_j} \mathbf{S}_i \cdot \mathbf{S}_{i'} u_m \eta_j + \text{terms of third and higher orders.} \quad (2)$$

Here, $E_H^0$ is the zero-order term with $u_m = 0$ and $\eta_j = 0$ [12], $J_{ii'}$ is the symmetric exchange interaction parameter between spins $\mathbf{S}_i$ and $\mathbf{S}_{i'}$, and $\frac{\partial J_{ii'}}{\partial u_m}$, $\frac{\partial J_{ii'}}{\partial \eta_j}$, $\frac{\partial^2 J_{ii'}}{\partial u_m \partial u_n}$, $\frac{\partial^2 J_{ii'}}{\partial \eta_j \partial \eta_k}$, $\frac{\partial^2 J_{ii'}}{\partial u_m \partial \eta_j}$ are the derivatives of the exchange parameters. Similarly, we can derive the expressions for $E_{DM}$ and $E_{SIA}$.

To obtain the structural distortion and cell deformation caused by the spin order, we can minimize the total energy E($u_m$, $\eta_j$, $S_i$) with respect to $u_m$ and $\eta_j$. Since $\frac{\partial^2 J_{ii'}}{\partial u_m \partial u_n} \ll B_{mn}$, $\frac{\partial^2 J_{ii'}}{\partial \eta_j \partial \eta_k} \ll B_{jk}$ and $\frac{\partial^2 J_{ii'}}{\partial u_m \partial \eta_j} \ll B_{mj}$, we finally obtain that

$$B_{mn} u_n + B_{mj} \eta_j = -\sum_{i,i'} \frac{\partial J_{ii'}}{\partial u_m} S_i \cdot S_{i'}$$

$$B_{mj} u_m + B_{jk} \eta_k = -\sum_{i,i'} \frac{\partial J_{ii'}}{\partial \eta_j} S_i \cdot S_{i'} \qquad (3)$$

By solving the above linear equations, we get the displacements $u_m$ and strains $\eta_j$. The spin-order induced strain can be used to obtain the new cell vectors $\mathbf{a}^{new}$: $\left[\mathbf{a}_1^{new}, \mathbf{a}_2^{new}, \mathbf{a}_3^{new}\right] = (I + \varepsilon)\left[\mathbf{a}_1^{PM}, \mathbf{a}_2^{PM}, \mathbf{a}_3^{PM}\right]$, where $\mathbf{a}^{PM}$ are the cell vectors of the PM state, I is a $3 \times 3$ unit matrix, and $\varepsilon$ is the strain matrix defined by $\eta_j$.

The magnetoelastic phenomena are associated with the dependence of the crystal cell vectors on the spin configurations. Using our above model, one can not only quantitatively compute the lattice change, but also reveal the microscopic origin of the interesting phenomena in great details. In particular, one can tell which spin site, spin pair, and type of the spin interaction are responsible for the magnetoelastic coupling. This is different from previous studies [13,14] in which the final macroscopic lattice response was obtained by changing the overall magnetic configuration of the system in the DFT calculations. In principle, we can use Eq. (3) to understand the magnetoelastic phenomena such as spin-order related NTE, magnetic phase transition induced lattice deformation, and magnetostriction. In the following of this work, we will show instead that the magnetoelastic coupling will give rise to a new contribution to the electric polarization induced by the spin-order, in which case the dimension of Eq. (3) may be greatly reduced.

Previously, it was shown [19-25] that spin-order induced electric polarization contains a pure electronic contribution and an ion-displacement related contribution (see Fig. 1). As we discussed above, spin-order may induce not only ion-displacement, but also lattice deformation. If the system in the PM state is piezoelectric (e.g., polar), we find that the lattice deformation induced by spin order may give rise to an additional electric polarization. Therefore, there is a lattice deformation contribution (see Fig. 1) to the electric polarization due to the combined effect of spin-order induced stress and piezoelectricity [16,18] in a magnetic material which belongs to one of the piezoelectric crystal classes in the PM state. In terms of $u_m$ and $\eta_j$, the polarization [26] can be computed as $P_\alpha = Z_{\alpha m} u_m + e_{\alpha j} \eta_j$, where $Z_{\alpha m}$ and $e_{\alpha j}$ are the Born effective charge and frozen-ion piezoelectric tensor, respectively. Here, both the ion-displacement and lattice deformation contributions are included in $P_\alpha$. Setting $-\sum_{i,i'} \frac{\partial J_{ii'}}{\partial u_m} \mathbf{S}_i \cdot \mathbf{S}_{i'} = 0$ in Eq. (3), one can obtain the polarization contribution due to the stress induced by spin-order. One can also evaluate this polarization contribution through the piezoelectric constant ($d_{\alpha j}$) by using $P_\alpha = \sum_j \sigma_j d_{\alpha j}$ where $\sigma_j = -\sum_{i,i'} \frac{\partial J_{ii'}}{\partial \eta_j} \mathbf{S}_i \cdot \mathbf{S}_{i'}$ is the total stress due to the spin order. And $d_{\alpha j}$ can be written as $d_{\alpha j} = S_{jk} e_{\alpha k}$ in which $e_{\alpha k}$ is the relaxed-ion piezoelectric tensor and $S_{jk}$ is the relaxed-ion elastic compliance tensor. Previously, Wojdel and Íñiguez [17] investigated the linear magnetoelectric (ME) coupling by including the piezoelectricity and piezomagnetism in BiFeO$_3$ and related materials. Their model can describe the overall linear ME coupling for the spin ground state. In this work, our model is generalized to include the spin interaction energy changes under different magnetic orderings and is able to describe higher-order (e.g., quadratic) ME coupling. Moreover, current model can also identify the exchange paths resulting in the particular magnetoelsatic coupling.

We will now discuss how to obtain the parameters in Eq. (3) within the first-principles framework. Density functional perturbation theory can be used to compute the force constant ($B_{mn}$), the internal-displacement tensor ($B_{mj}$). The frozen-ion elastic constant ($B_{jk}$) can be easily obtained by calculating the strain-stress relation within DFT. To compute the first-order derivatives of the symmetric spin exchange parameter $J_{ii'}$ with respect to $\eta_j$, we propose a four-states mapping approach: $\frac{\partial J_{ii'}}{\partial \eta_j} = \frac{1}{4}(\frac{\partial E_I}{\partial \eta_j} + \frac{\partial E_{IV}}{\partial \eta_j} - \frac{\partial E_{II}}{\partial \eta_j} - \frac{\partial E_{III}}{\partial \eta_j}) = -\frac{1}{4}(\sigma_j^I + \sigma_j^{IV} - \sigma_j^{II} - \sigma_j^{III})$ (see Fig. 2). Here, I-IV refer to the four spin states with different spin orientations for sites i and i' (see Fig. 2 for an example), E and $\sigma$ denote the total energy and stress, respectively. We note that the stress can be computed without doing extra DFT calculations due to the celebrated Hellmann-Feynman theorem. The first-order derivatives of the symmetric spin exchange parameter $J_{ii'}$ with respect to $u_m$ can be also efficiently evaluated by using a four-states mapping approach [12].

In the following, we will apply our general model of magnetoelastic coupling to the classic room-temperature multiferroic BiFeO$_3$. BiFeO$_3$ [27-29] crystallizes in a *R*3*c* structure with a large polarization ($\sim 100\,\mu C/cm^2$) [30] when the temperature is lower than the FE Curie temperature $T_C$ = 1000K. On cooling below $T_N$ = 650K, a G-type AFM order with a long period incommensurate modulation takes place. Interestingly, some experiments [31-33] discovered the ME coupling in BiFeO$_3$. However, how magnetoelectric coupling actually occurs on a microscopic level in multiferroic BiFeO$_3$ is not clear. We will investigate the microscopic origin of the ME coupling in BiFeO$_3$ from our model. Our total energy calculations are based on the DFT plus the on-site repulsion (U) method [34] within the generalized gradient approximation [35] (DFT+U) on the basis of the projector augmented wave method [36] encoded in the Vienna ab initio simulation package (VASP) [37]. The plane-wave cutoff energy is set to 500 eV in the DFT calculations unless noted otherwise. The on-site repulsion U and exchange parameter J are set to 5 and 1 eV for Fe. For the calculation of electric polarization, the Berry phase method [38] is used.

Our new four-states approach for computing $\frac{\partial J_{ii'}}{\partial \eta_j}$ is compared with the a conventional finite difference method in which the exchange interactions at different strains are computed explicitly.

To compute all $\frac{\partial J_{ii'}}{\partial \eta_j}$ (j = 1-6) for a given exchange interaction $J_{ii'}$, the finite difference method requires 48 DFT total energy calculations, while only 4 total energy calculations are needed in the four-states approach. Thus, the four-states approach is computationally more efficient and convenient. To check the accuracy of the four-states approach, we take BiFeO$_3$ as an example. A $2\times2\times2$ supercell of a rhombohedra *R3c* structure is adopted to compute $\frac{\partial J_{NN}}{\partial \eta_j}$, where $J_{NN}$ is the nearest neighboring (NN) Fe-Fe spin exchange interaction in BiFeO$_3$. The plane-wave cutoff energy is increased to 700 eV in order to obtain converged results for the stress. The results are presented in Table I. Our subsequent analysis shows that $\frac{\partial J_{NN}}{\partial \eta_3}$ plays the most important role on the magnetoelastic coupling in BiFeO$_3$. Therefore, we also use the finite difference method to evaluate $\frac{\partial J_{NN}}{\partial \eta_3}$ in which $J_{NN}$ is calculated as a function of the strain ($\eta_3$) ranging from 0 to 0.006. As shown in Fig. 3(a), the plot of $J_{NN}$ versus $\eta_3$ is a straight line in the studied region, thus we can obtain $\frac{\partial J_{NN}}{\partial \eta_3} = -0.088$ eV that is very close to that ($-0.084$ eV) obtained from our four-states approach.

Our above calculations show that $\frac{\partial J_{NN}}{\partial \eta_3}$ is negative, i.e., a positive strain along the z-axis makes $J_{NN}$ smaller. We will understand the dependence of $J_{NN}$ on $\eta_3$ on the basis of the superexchange theory. As shown in Fig. 3(b), when $\eta_3$ is positive, the $\angle$Fe1-O-Fe2 angle ($\theta$) will become closer to 180° and the Fe1-O and Fe2-O bond lengths will be elongated. According to the Goodenough-Kanamori rule, the superexchange interaction J is proportional to $\frac{t^2}{U}$ [39,40], where t and U are the effective orbital hopping and Hubbard repulsion, respectively. A larger angle makes the hopping stronger, while the longer bond length weakens the hopping. Therefore, this qualitative analysis is not able to determine how $J_{NN}$ will change. Quantitatively speaking, the effective hopping between the 3d orbitals of Fe1 and Fe2 can be approximately expressed as $t = t_1^{pd\sigma} t_2^{pd\sigma} \cos\theta$, where $t_i^{pd\sigma}$ is the hopping integral between the $e_g$ orbital of the i-th Fe ion and the 2p orbital of the intermediate O ion. Because $t_i^{pd\sigma}$ is proportional to $\frac{1}{|\mathbf{l}_i|^4}$ [the distance vector $\mathbf{l}_i$ is defined in Fig. 3(b)] [41], we find $t \sim \frac{\cos\theta}{|\mathbf{l}_1|^4 |\mathbf{l}_2|^4}$. Expanding $|\mathbf{l}_i|$ and $\cos\theta$ as a function of $\eta_3$, we obtain $t \sim \frac{\mathbf{l}_{10} \cdot \mathbf{l}_{20} + \alpha \eta_3}{|\mathbf{l}_{10}|^5 |\mathbf{l}_{20}|^5}$, where $\mathbf{l}_{i0}$ is the original distance vector with $\eta_3 = 0$, and $\alpha = 2|\mathbf{l}_{10}^z||\mathbf{l}_{20}^z| - 5\mathbf{l}_{10} \cdot \mathbf{l}_{20} [\frac{|\mathbf{l}_{10}^z|^2}{|\mathbf{l}_{10}|^2} + \frac{|\mathbf{l}_{20}^z|^2}{|\mathbf{l}_{20}|^2}]$. One can easily see [42] that $\alpha < 0$, thus t becomes smaller for a positive $\eta_3$ and $\frac{\partial J_{NN}}{\partial \eta_3} < 0$, in consistent with the DFT result. Similarly, we can demonstrate that $\frac{\partial J_{NN}}{\partial \eta_1} < 0$ and $\frac{\partial J_{NN}}{\partial \eta_2} < 0$.

From our model, we can compute the total stress resulting from the ordering of the G-type AFM order by using $\sigma_{AFM} = -\sum_{<ii'>_{NN}} \frac{\partial J_{ii'}}{\partial \eta_j} \mathbf{S}_i \cdot \mathbf{S}_{i'}$, where only the NN Fe-Fe pairs are considered. This stress can be compared to the direct DFT value from a DFT calculation on BiFeO$_3$ in the G-AFM spin state with the equilibrium structure of the PM state (simulated by two orthogonal spins in the 10-atom rhombohedra cell). Table I indicates a good agreement between the model and the direct DFT calculation. This also suggests that $\frac{\partial J_{NN}}{\partial \eta_j}$ is sufficient for describing the magnetoelastic coupling in BiFeO$_3$. We now turn to examine how the magnetoelastic coupling influences the electric polarization in BiFeO$_3$. By solving Eq. (3), we find that the strain is $\eta$ = (-8.26, -8.26, - 35.58, 0, 0, 0) in the order of $10^{-4}$ as a result of the G-AFM ordering. Mediated by the coupling between polarization and strain, the lattice change will induce a polarization. As can be seen in Table II, our model predicts a lattice deformation contribution to the polarization of P = 1.32 $\mu C/cm^2$, which is even larger than the sum of the pure electronic and ion-displacement contributions. This is an unprecedented result in that a previously unknown contribution to electric polarization induced by spin order is found to be even larger than the widely known contributions. Table II shows that the result obtained from our model is also in agreement with the direct DFT calculations. Summing up all the three spin-order induced contributions with the same sign, the total polarization calculated for the G-type AFM order in BFO reaches ~2 $\mu C/cm^2$. The spin-induced polarization in BFO is also comparable with that of HoMnO$_3$ [24,43]. We find that the direction of the polarization caused by the spin order is opposite to the inherent electric polarization due to the *R*3*c* structure distortion. This is consistent with a recent experimental observation [31]. In that experiment [31], the ion-displacement contribution deduced from the displacement of the Fe ions was determined to be 0.4 $\mu C/cm^2$, which is also close to the value (0.56 $\mu C/cm^2$) obtained from our model.

Some experiments [32,33] suggested that an external magnetic field may change the electric polarization of BiFeO$_3$. Qualitatively, we can understand the ME coupling in BiFeO$_3$ from our model. Considering only the NN spin exchange interaction and Zeeman term, the total energy can be written as $E = \sum_{\langle i,i'\rangle_{NN}} J_{NN} \mathbf{S}_i \cdot \mathbf{S}_{i'} - \mu_B g \sum_i \mathbf{S}_i \cdot \mathbf{H}$, where $\mu_B$, g and **H** are Bohr magneton, Landé factor and magnetic field, respectively. By minimizing the total energy, the angle $\theta$ between the two spins $\mathbf{S}_1$ and $\mathbf{S}_2$ in the 10-atom cell in a magnetic field is $\theta = 2\arccos(\frac{5\mu_B H}{12 J_{NN}})$ (the effective $J_{NN}$ = 35.76 meV in our study). As can be seen from Eq. (3), the spin-order induced polarization $P \propto \langle \mathbf{S}_i \cdot \mathbf{S}_{i'} \rangle \propto \cos\theta$. It can be easily shown that $\Delta P = P(H) - P(0) \propto H^2$. Therefore, we obtain a quadratic dependence of this spin-order induced polarization on the magnetic field, i.e., the quadratic ME coupling (see Fig. 4). At a magnetic field of 20 T, we find that $\Delta P = 9 \times 10^{-4} \, \mu C/cm^2$, which is in agreement with the result from one experiment [32], but there is a large discrepancy between our result and another experimental result [33]. Note that our above analysis is based on a simplified spin Hamiltonian without DM interactions and single-ion anisotropy. Further experimental and theoretical studies are called for to resolve this discrepancy.

In summary, we propose a microscopic model that describes magnetoelastic coupling. All the parameters in this model can be computed from first-principles. In particular, we propose an efficient four-states approach for computing the derivate of the spin interaction parameter with respect to the strain. On the basis of this model, we reveal that there exists a previously unexpected contribution to the electric polarization induced by the spin-order in multiferroics due to the combined effect of magnetoelastic coupling and piezoelectric effect. Interestingly, we find that this lattice deformation contribution to the polarization is even larger than that from the pure electronic and ionic contributions in $BiFeO_3$. The spin-order induced polarization is opposite to the proper polarization due to the R3c distortion, in agreement with the negative ME effect observed experimentally [31]. Furthermore, how an external magnetic field modulates the electronic polarization in $BiFeO_3$ is discussed qualitatively by using the general model. Our microscopic model of magnetoelastic coupling will be useful to investigate the linear and higher order ME effects and the origin of magnetoelastic phenomena.

$2|\mathbf{l}_{10}^z||\mathbf{l}_{20}^z|+5(|\mathbf{l}_{10}^z|^2+|\mathbf{l}_{20}^z|^2)\cos\theta$ where $\theta=153°$. Because $2|\mathbf{l}_{10}^z||\mathbf{l}_{20}^z|\leq|\mathbf{l}_{10}^z|^2+|\mathbf{l}_{20}^z|^2$, $\alpha$ should be a negative value.

**Acknowledgements**

X. Wu was supported as part of the Center for the Computational Design of Functional Layered Materials, an Energy Frontier Research Center funded by the U.S. Department of Energy, Office of Science, Basic Energy Sciences under Award #.DE-SC0012575. Work at Fudan was supported by NSFC, the Special Funds for Major State Basic Research, NCET-10-0351, Research Program of Shanghai Municipality and MOE, Program for Professor of Special Appointment (Eastern Scholar), and Fok Ying Tung Education Foundation.

Table I. First-order derivative of the nearest-neighbor (NN) spin exchange parameter with respect to the strain $\eta_j$ ($\frac{\partial J_{NN}}{\partial \eta_j}$) computed by using the four-states approach. The total stress ($\sigma_j$) induced by the G-type AFM order in BiFeO$_3$ from the model and DFT calculations is presented as well.

| J | 1 | 2 | 3 | 4 | 5 | 6 |
|---|---|---|---|---|---|---|
| $\frac{\partial J_{NN}}{\partial \eta_j}$ (eV) | -0.086 | -0.041 | -0.084 | 0.022 | 0.075 | -0.029 |
| $\sigma_j$ (kB) Model | -4.769 | -4.769 | -6.322 | 0 | 0 | 0 |
| $\sigma_j$ (kB) DFT | -4.420 | -4.420 | -5.475 | 0 | 0 | 0 |

Table II. The different contributions to the electric polarization (in unit of $\mu C/cm^2$) induced by the G-AFM order in BiFeO$_3$ from model and DFT calculations. P$_{lattice}$, P$_e$ and P$_{ion}$ refer to the lattice deformation, pure electronic and ion displacement contributions, respectively.

| Polarization | P$_{lattice}$ | P$_e$ | P$_{ion}$ |
|---|---|---|---|
| Model | 1.32 | 0.53 | 0.56 |
| DFT | 1.22 | 0.40 | 0.54 |

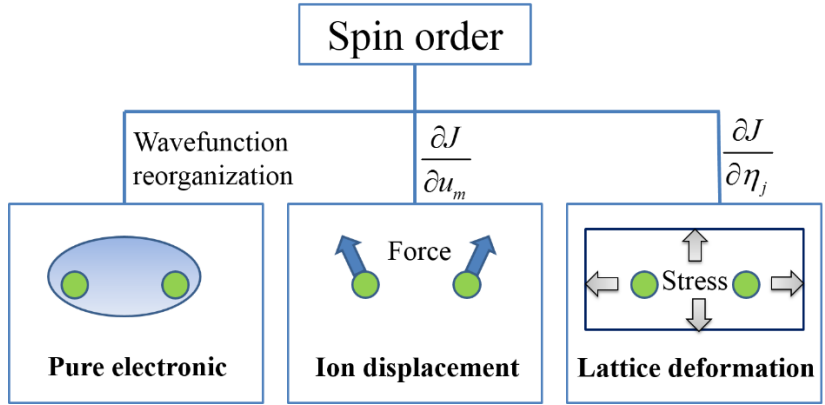

Figure 1. Schematic illustration of three contributions to the electric polarization induced by a spin-order in multiferroics. The pure electronic contribution [19,21,22] arises from the electron density redistribution induced by the spin-order. For the ion-displacement part, it results from the ion displacements caused by the induced forces associated with a spin order [20,24]. In this work, we reveal a new contribution, i.e., the lattice-deformation contribution, which results from the spin-order induced stress (i.e., the magnetoelastic coupling).

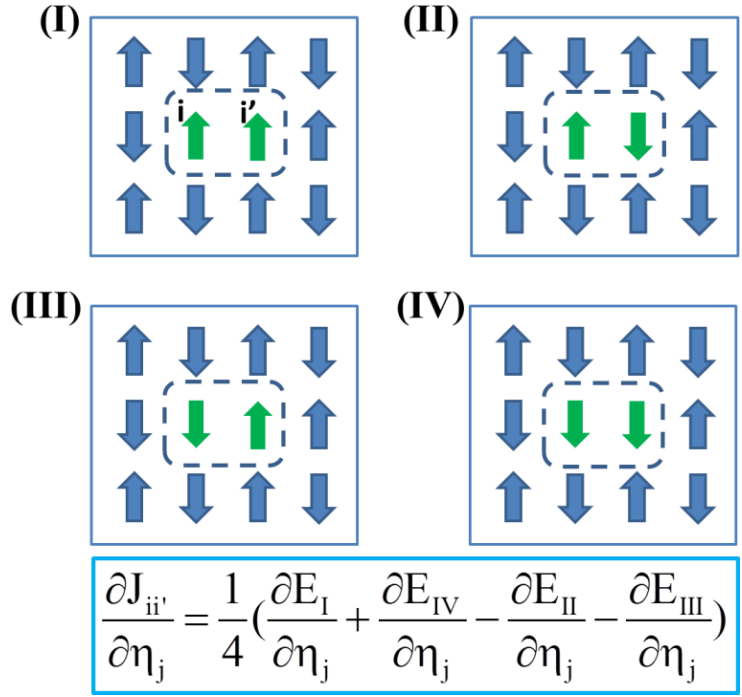

Figure 2. Schematic illustration of the four spin states in the four-states approach to calculate the derivative of exchange parameter with respect to strain $\frac{\partial J_{ii'}}{\partial \eta_j}$. In the four spin states, only the spins at sites i and i' change the orientation.

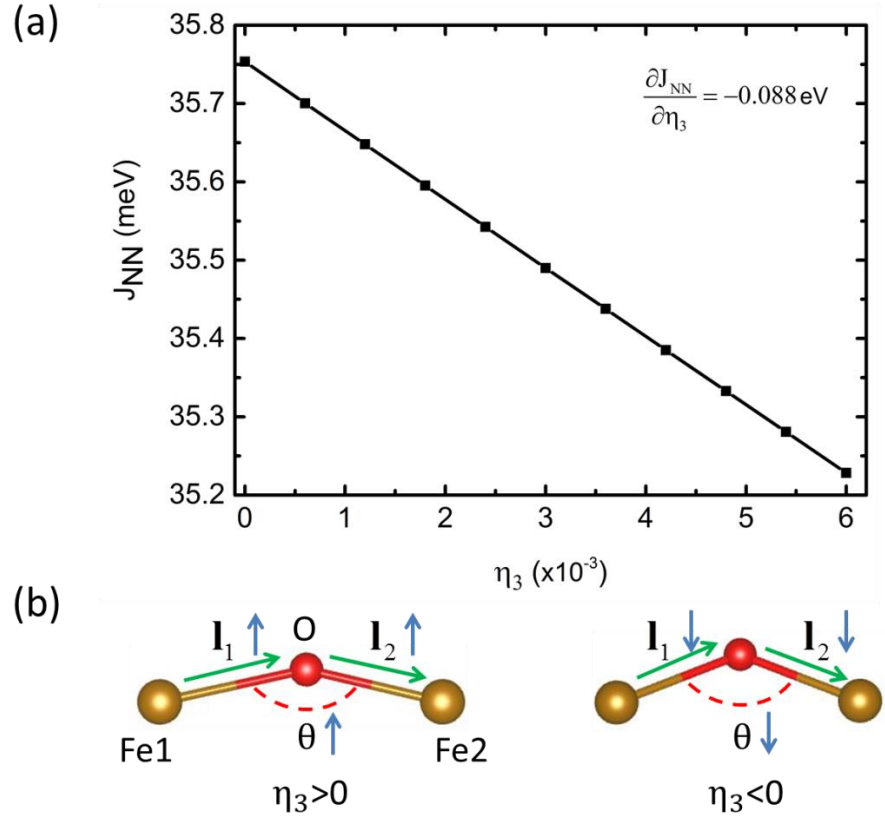

Figure 3. (a) The NN symmetric spin exchange interaction $J_{NN}$ as a function of $\eta_3$. The obtained $\frac{\partial J_{NN}}{\partial \eta_3}$ from the finite difference method is in good agreement with that ($\frac{\partial J_{NN}}{\partial \eta_3}$ =-0.084 eV) from the four-states approach. (b) Illustrations of the changes of bond lengths ($|\mathbf{l}_1|$, $|\mathbf{l}_2|$) and angle ($\theta$) with strain ($\eta_3$) in a Fe1-O-Fe2 system related to $J_{NN}$. Green arrows indicate the directions of $\mathbf{l}_1$ and $\mathbf{l}_2$.

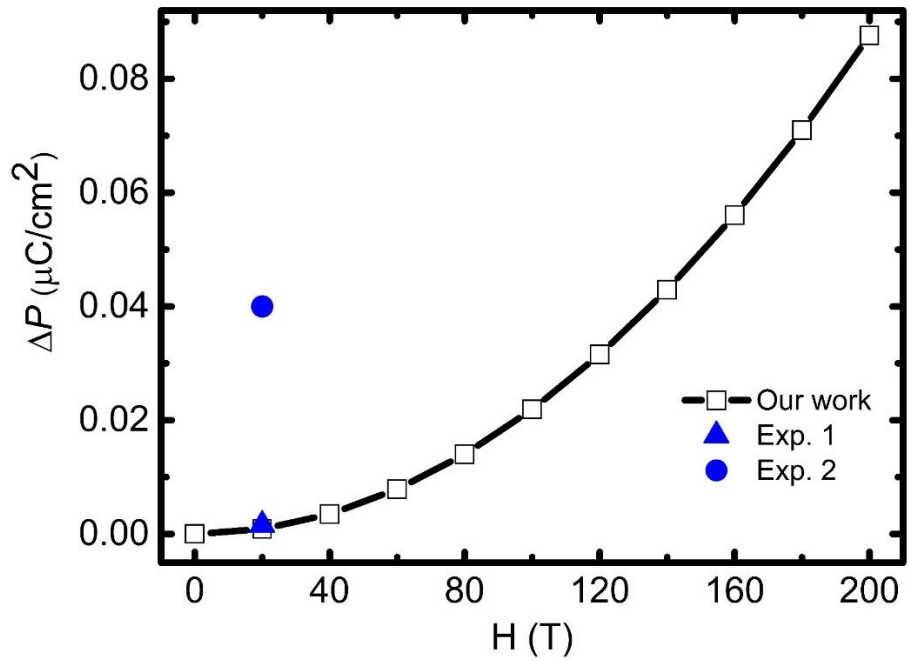

Figure 4. Polarization (P) versus magnetic field (H) calculated from our simple theoretical model. $\Delta P$ is defined as $\Delta P = P(H) - P(0)$. Experimental results (Exp.1 [32] and Exp. 2 [33]) are also shown for comparison.